\documentclass{WileyMSP-template}

\usepackage{graphicx}
\usepackage{graphics}
\usepackage{amssymb}
\usepackage{amsmath}
\usepackage{epsfig}
\usepackage{latexsym}
\usepackage{color}
\usepackage{rotating}
\usepackage{subfigure}
\usepackage{float}
\usepackage[breaklinks=true]{hyperref}
\usepackage{siunitx}
\usepackage[linewidth=1pt]{mdframed}
\usepackage{lipsum}
\usepackage{setspace}
\usepackage{comment}
\usepackage{ragged2e}

\hypersetup{
  colorlinks   = true, %Colours links instead of ugly boxes
  urlcolor     = blue, %Colour for external hyperlinks
  linkcolor    = blue, %Colour of internal links
  citecolor   =  red %Colour of citations
}

\usepackage{soul} % for strikethrough \st

\begin{document}

\pagestyle{fancy}
\rhead{\includegraphics[width=2.5cm]{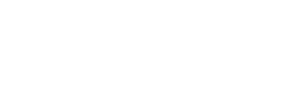}}

\title{Quantum Biotechnology}

\maketitle

% Author: Please give full first and last names for authors and include * after the name of all corresponding authors

\author{Nicolas P. Mauranyapin}
\author{Alex Terrason}
\author{Warwick P. Bowen*}

% Dedication
\dedication{}

% Affiliations: Please provide adacemic titles (Prof. or Dr.) for all authors where applicable, and include an institutional email address for all corresponding authors
\begin{affiliations}
Nicolas Mauranyapin, Alex Terrason and Warwick P. Bowen\\
ARC Centre of Excellence for Engineered Quantum Systems, University of Queensland, St Lucia, QLD 4072, Australia\\
Email Address: w.bowen@uq.edu.au

\end{affiliations}

% Keywords: Please provide a minimum of three and a maximum of seven keywords, separated by commas


\justifying

\begin{abstract}
Quantum technologies leverage the laws of quantum physics to achieve performance advantages in applications ranging from computing to communications and sensing. They have been proposed to have a range of applications in biological science. This includes better microscopes and biosensors, improved simulations of molecular processes, and new capabilities to control the behaviour of biomolecules and chemical reactions. Quantum effects are also predicted, with much debate, to have functional benefits in biology, for instance, allowing more efficient energy transport and improving the rate of enzyme catalysis. Conversely, the robustness of biological systems to disorder from their environment has led to proposals to use them as components within quantum technologies, for instance as light sources for quantum communication systems. Together, this breadth of prospective applications at the interface of quantum and biological science suggests that quantum physics will play an important role in stimulating future biotechnological advances. This review aims to provide an overview of this emerging field of \textit{quantum biotechnology}, introducing current capabilities, future prospects, and potential areas of impact. The review is written to be accessible to the non-expert and focuses on the four key areas of quantum-enabled sensing, quantum-enabled imaging, quantum biomolecular control, and quantum effects in biology.

\end{abstract}

\section{Introduction}

Biotechnology has transformed human society, from more effective treatment of disease, to better agricultural productivity, and improved clean energy generation and storage~\cite{khan2020biotechnology}. Broadly defined, the field of biotechnology seeks to understand biological processes and to exploit them in areas ranging from genomics and proteomics, to pharmaceuticals and immunology~\cite{khan2020biotechnology}.
Advances in the technologies used to observe, understand and engineer nanoscale biological systems have been a key driver of progress, underpinning capabilities such as gene sequencing~\cite{logsdon2020long, lappalainen2019genomic} and gene editing~\cite{doudna2020promise,yin2017progress}, and enabling  improved design of drugs, chemicals, soft materials and artificial biomaterials~\cite{friedman2020engineering}. Quantum technologies are a rapidly evolving class of technologies that are expected to  significantly contribute to the future of this technology-driven progress.

Quantum technologies have been identified to offer powerful prospective applications in computing and sensing~\cite{thew2019focus, riedel2017european}. Their attraction is their ability to achieve performance that goes beyond the fundamental limits of other technologies. Quantum computers allow some information processing tasks to be performed with speed that is exponentially faster than conventional computers~\cite{arute2019quantum}, quantum control techniques give access to fundamentally new types of behaviour~\cite{wiseman2009quantum}, quantum imaging technologies can resolve fainter and smaller structures than is possible with other techniques~\cite{Taylor2016},
while quantum sensors allow exquisitely precise nanoscale measurements of magnetic and electric fields, spins, and temperature~\cite{degen2017quantum}. This review aims to present a forward-looking and accessible perspective on the applications of these quantum technologies in biotechnology.

\section{Overview}

Magnetic resonance imaging (MRI) and nuclear magnetic resonance (NMR) are perhaps the most widely known examples of quantum technology applied into biology today. Here, strong magnetic fields are used to polarise and drive the dynamics of the nuclear spins in a specimen, with the response providing an image or chemically specific information~\cite{brown2014magnetic,levitt2013spin}. Rapid progress is being made to miniaturise these magnetic resonance systems so that they can be applied to single atoms and molecules, and so that deeply quantum effects such as quantum entanglement may be exploited within them~\cite{degen2017quantum}. A variety of other quantum nanosensors and quantum microscopes are also being developed, allowing us to see further into biological samples and to see them at greater detail~\cite{Taylor2016,degen2017quantum,levine2019principles}. For instance, quantum technologies provide new ways to break the diffraction limit of imaging~\cite{tenne2019super}, allow improved imaging contrast within a constrained photon budget~\cite{Casacio2021}, and enable nanoscale measurements of intracellular dynamics~\cite{barry2016optical,le2013optical,taylor2013biological}.

Extending beyond improved sensing and imaging technologies, quantum technologies  offer new ways to control and simulate biological systems. Quantum control can be achieved, for instance, by confining a biomolecule or ensemble of molecules within a nanoscale optical cavity~\cite{herrera2020molecular,sanvitto2016road}.
This can be used to drive chemical reactions that would otherwise not occur~\cite{keefer2018pathways,shapiro2012quantum}, and to create new types of matter -- hybrid polaritonic states that share optical and molecular properties. Such control could be used to transfer quantum states between light and biomolecular systems~\cite{chikkaraddy2016single}, and therefore to form the basis of a class of robust room temperature technologies for quantum computing, communication and sensing~\cite{herrera2020molecular,chikkaraddy2016single}. Quantum computers allow molecular simulations that go beyond what is possible with quantum chemistry~\cite{aaronson2009quantum, elfving2020will}. These fully-quantum simulations are predicted to allow faster, more accurate simulations of large molecules~\cite{elfving2020will, arguello2019analogue}, and could therefore play an important role in the design of future pharmaceuticals, artificial enzymes and energy harvesting systems.

Quantum biology is a further active, and sometimes controversial, area of research that resides at the interface of quantum and bio-science. Here, rather than designing quantum technologies to {\it learn} about biology, the aim is to define the extent to which quantum effects play a functional role {\it in} biology~\cite{lambert2013quantum}. It has been suggested with much debate that quantum effects are important for processes such as photosynthesis, enzyme catalysis and neural signaling~\cite{kim2021quantum,marais2018future}. If this is the case, it could have profound consequences for our understanding of biology and our ability to design functional biological systems. Ongoing efforts are underway to find definitive answers. As quantum technologies to observe, simulate and control biological systems advance, it can be anticipated that they will greatly augment these efforts.

The four areas described above, of quantum  sensing and imaging, quantum molecular control, quantum chemical simulation, and quantum biology, form the main focus of this review.
By bringing them together into one review, we hope to 
provide a coherent perspective on the breadth of impact that quantum technologies may have on biotechnology. We also hope to illustrate the complementarity of the advances being made within the different areas -- for example, better molecular sensors and better simulations together could provide deeper understanding of molecular dynamics and of whether quantum effects play a functional role within them.

\section{Quantum-enabled sensing}

%Introduction
Sensors are essential components of our daily life. They are used to detect fluctuations of a variety of physical quantities such as temperature, electric and magnetic fields as well as stress and strain, and to detect biological samples including fragments of tissue, microorganism, single cells, single molecules and small proteins. This enables a wide range of biotechnological applications, ranging from biomedical diagnostics to proteomics, genomics, chemistry, pharmaceutical production, and environment and agricultural monitoring \cite{teles2011biosensors,karlsson2004spr,yao2014chemistry,ahmed2014personalized,verma2015biosensor,velasco2003biosensor}. A significant underpinning application is their use to observe molecular dynamics, which has foundational importance for our understanding of  protein-protein interactions~\cite{kengne2012protein}, how diseases develop~\cite{haleem2021biosensors}, the efficiency of treatments~\cite{haleem2021biosensors}, and important industrial processes such as catalysis and biological energy transport~\cite{mukundan2004electrochemical}.

The broad biotechnological applications of biosensors have driven rapid development, focused on enhancing sensitivity and resolution, decreasing the time required for detection, and targeting different biologically-relevant stimuli. 
Quantum technologies offer new tools and approaches to improve these characteristics~\cite{degen2017quantum}. There are several ways to leverage quantum effects. For example, 
through the detection of quantum processes either inherent in the biological sample itself, or within a quantum probe introduced into it~\cite{gunther2013nmr, brown2014magnetic,levitt2013spin,staudacher2013nuclear,barry2016optical}, or through the use of quantum coherences/entanglement to improve signal and noise characteristics compared to classical systems~\cite{taylor2013biological,taylor2014subdiffraction,crespi2012measuring,dowran2018quantum}. In this section, we will give examples of these quantum biosensing techniques and describe existing and future applications within biotechnology.

%NMR
We begin with the example of NMR spectroscopy which, as discussed in the previous section, is a widely known quantum sensing technique. It was discovered in the 1940s and probes the nuclear spins of biochemical samples~\cite{gunther2013nmr}. During NMR measurements, a sample in solution is exposed to a large static magnetic field. In this configuration, due to the Zeeman effect, the nuclei magnetic dipole (spin) energy states of NMR-active atoms in the sample (typically isotopes 1H, 13C, 15N and 31P) split~\cite{gunther2013nmr}. Transitions between these energy states can then be resonantly driven by an additional modulated magnetic field.
The response of the magnetic spin state to this driving can be measured with sensitive magnetic field sensors, and provides information about both the chemical content of the sample and its molecular structure~\cite{brown2014magnetic,levitt2013spin}.

An accurate understanding of molecular structures is crucial for biotechnology, since structure is a key determining factor in molecular function and interactions. Here, NMR spectroscopy is a crucial tool, combined with X-ray crystallography, electron microscopy and molecular simulations~\cite{dill2012protein}. Indeed, NMR is, to date, the most powerful tool for characterizing the structures of organic and biochemical compounds in solution~\cite{gunther2013nmr}. For example, the  protein data bank includes more than 13,000 NMR-derived three-dimensional protein structures~\cite{pdb}.

NMR can also be used to measure other molecular properties such as the magnitude of the angular fluctuations of chemical bond vectors, providing information about protein dynamics~\cite{ishima2000protein} as well as backbone dynamics involved in protein-protein and protein-nucleic acid binding~\cite{wikstrom1999conformational, huang1999backbone}.
Together, these capabilities offer a multitude biotechnological applications. This is especially the case in medicine, where NMR spectroscopy has been successfully used in diagnostics, blood flow measurements, metabonomic studies and development of drug delivery, among many others~\cite{damadian2013nmr}. NMR is also applied widely in other areas such as food sciences~\cite{hatzakis2019nuclear} and geophysics~\cite{behroozmand2015review}, with the recent development of portable NMR systems advancing capabilities in geological surveying and medical point-of-care applications~\cite{dupre2019micro}.

While NMR is a powerful and well established technique, as described above, it suffers significant drawbacks in molecular studies. NMR signals are inherently weak~\cite{xu2018molecular}, necessitating repeated measurements and averaging. As a result, the speed is far slower than the characteristic timescales of most molecular processes. To achieve sufficient signal-to-noise, the technique also generally requires collective measurements on large ensembles of perhaps a trillion molecules,
and therefore averages away important heterogeneity and single-molecule dynamics~\cite{xu2018molecular}.

The ability to study single molecules, and ideally resolve their dynamics in real-time, is
important to shed light on the behaviour of proteins, which are responsible for most microscopic biological  processes, including catalysing and regulating biochemical reactions, production and transcription of other proteins, photosynthesis, muscular motion, signal transduction and transport of material notably in the brain~\cite{berg2002biochemistry}.
Proteins are not static but rather continuously in motion, transitioning between conformational states which depend on their three dimensional structure. How proteins fold and unfold between these states is one of the fundamental problems in biophysical science and has been studied for more than half a century~\cite{dill2012protein,dill2008protein,chan1993protein}.
Single molecule studies are challenging since proteins can be as small as few hundreds of Dalton and transitions between states can occur on time scales down to nanoseconds~\cite{anunciado2017vivo}. Although, techniques that utilise chemosensing methods and nanosized detectors have been developed to improve NMR systems~\cite{xu2018molecular}, these techniques have sensitivities very far away from what is needed to resolve single proteins and their dynamics. These limitations are being addressed by exploiting quantum nanoprobes.

%Quantum nanoprobes
Quantum nanoprobes are nanoscale quantum systems that are sufficiently robust to be applied in a biological environment and can be used to label single molecules and biosamples. One of the principal quantum nanoprobes used in biosensing are nanodiamond probes, which we will focus on here. Compared to other quantum probes such as quantum dots, they are non-toxic~\cite{mochalin2012properties} and their surface chemistry allows binding of a variety of biomolecules~\cite{cheng2008facile,mochalin2011covalent}. Nanodiamonds can be fabricated using many different processes~\cite{danilenko2004history} including detonation, laser ablation~\cite{yang1998preparation} and high-energy ball milling of high-pressure high-temperature diamond microcrystals~\cite{boudou2009high}. Their quantum properties arise from the electron spin of nitrogen vacancy (NV) centres which can be created through irradiation of high-energy electrons or helium atoms~\cite{su2013creation}.
Negatively charged NV-centers have long spin coherence times (up to miliseconds~\cite{balasubramanian2009ultralong} at room temperature). Their electron spin energy levels allow remote optical initialisation as well as optical and microwave read-out~\cite{balasubramanian2014nitrogen,mochalin2012properties}.

Single NV-centers in diamond plates have been used to perform nanometer-cubed NMR measurements in solution~\cite{staudacher2013nuclear,mamin2013nanoscale}. Since the magnetic field created by a spin scales with the inverse of the distance from the spin cubed, the ability to place a NV-center in close proximity to the targeted spin allows greatly enhanced sensitivity. Together with the ability to optically initialise and read-out the NV-centre at room temperature, this has enabled the measurement resolution to reach the level of single electronic and nuclear spins~\cite{grinolds2013nanoscale,neumann2010single}, far outperforming conventional NMR systems.

The sensitivity of conventional NMR is not only limited due to their large size scale,  but also from the low polarisation of nuclear spin-states at room temperature. Under typically accessible magnetic fields, the spacing of the nuclear spin energy levels is much smaller than the characteristic thermal energy $k_B T$, where $k_B$ is Boltzmann's constant and $T$ is the temperature. As a result, the upper energy level has nearly identical population to the lower level (typically less than 0.1 percent reduced at room temperature even for high DC magnetic fields~\cite{meier2014hyperpolarized}), and therefore the NMR contrast is very low. 
The optical initialisation of NV centres allows them to be prepared in a high polarisation state, with  the upper level population suppressed close to zero~\cite{witte2013nmr}.  They can therefore be used directly as hyperpolarised probes~\cite{bucher2020hyperpolarization,eills2019high}. Their high polarization can  also be transferred to nuclear spins of the sample~\cite{schwartz2019blueprint}.
This results in a drastically improvement of the polarizability of the sample, boosting the sensitivity of NMR measurements by up to five orders of magnitudes~\cite{ardenkjaer2003increase,meier2014hyperpolarized}. 
Other probes, such as noble gases for example, can also be hyperpolarised using techniques such as para-Hydrogen-induced polarisation, spin exchange optical pumping and the most popular dynamic nuclear polarisation~\cite{meier2014hyperpolarized,witte2013nmr} and applied prior to the NMR measurement to similarly increase the sensitivity of the measurement.

\begin{figure}[!ht]
 \centering
 \includegraphics[width=1\linewidth]{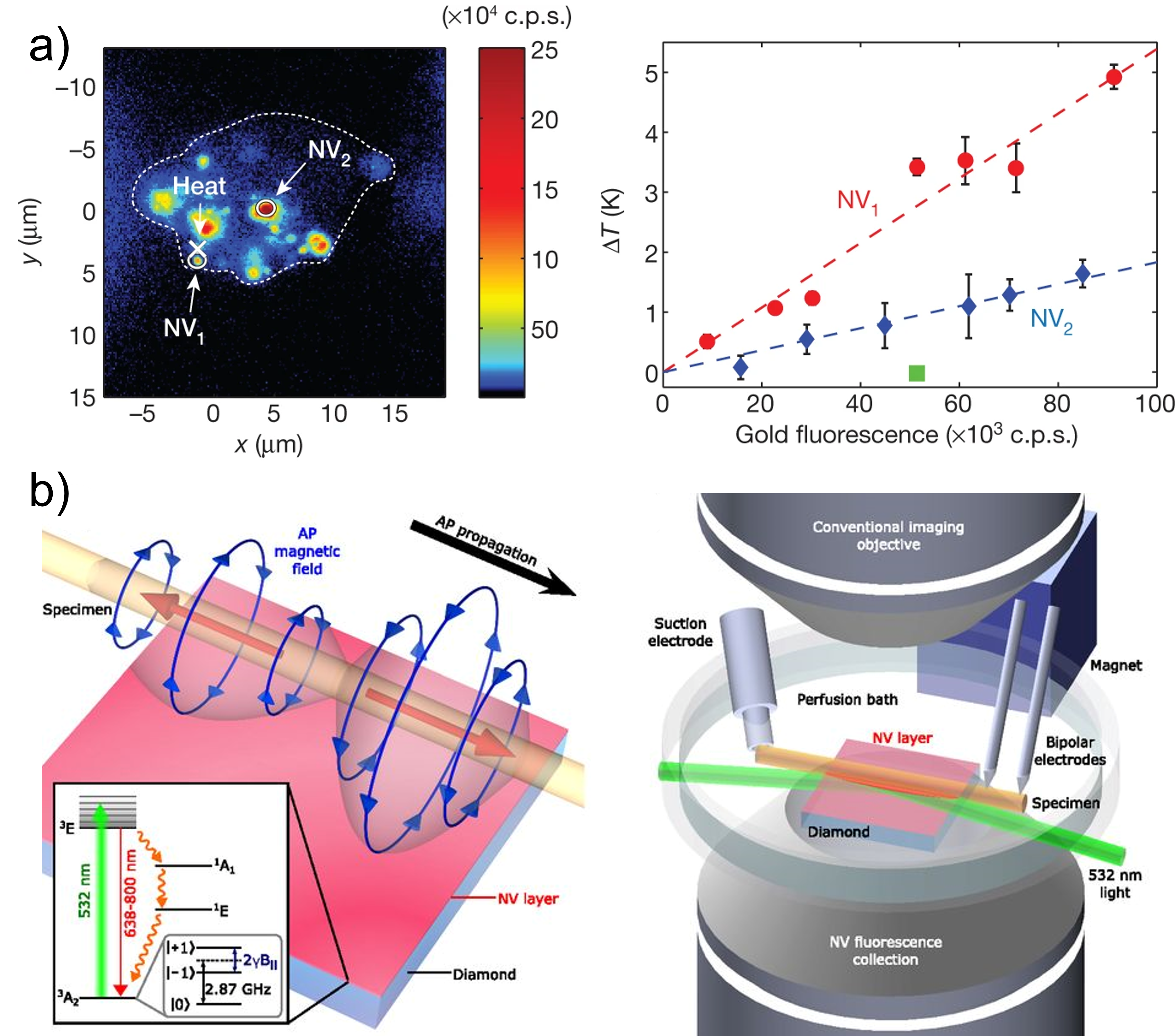}
 \caption{Biosensing with nanodiamonds. a) Temperature sensing inside a single cell using two nanodiamond NV-centers. Left, confocal scan showing the cell studied and the two NV-center used for temperature sensing. Right, measured temperature change of the two NV-centres as function of optical power applied to heating gold particle inside the cell  (Reproduced from Ref.~\cite{kucsko2013nanometre} with permission 5175770283468). b) Detection of action potential using NV-centers inside the neuron of a living worm. Left, experimental setup showing a diamond slab in blue with a layer of NV-center in red on which the axon of the specimen rests (orange). Right, the NV-centers are excited with green light and their red fluorescence is collected by the bottom objective while the top objective is used for imaging (Reproduced with permission from Ref.~\cite{barry2016optical}, Copyright (2016) National Academy of Sciences).  } 
\label{fig:NV}
\end{figure}

The quantum properties of NV-centres can also be used to extract other information from nanoscale samples with extreme precision. NV-centres can fluoresce  with extreme stability compared to other labels and cannot photobleach~\cite{balasubramanian2014nitrogen}. This makes them an attractive solution for fluorescence imaging (see following section). The fluorescence rate of NV-centres depends on their spin states, allowing spin state detection using optically detected magnetic resonance~\cite{balasubramanian2014nitrogen,mochalin2012properties}. 
Since the energy difference between the $m=0$ and $m=\pm1$ ground states has a temperature or strain dependence, nanodiamonds can be used to measure these quantities in biosamples at nanoscale. Temperature variation down to 1.8 mK have been observed as well as temperature-gradient control and mapping in human embryonic fibroblast (see figure \ref{fig:NV}a). This has biotechnology applications in temperature-induced control of gene expression, tumour metabolism and cell-selective treatment of disease~\cite{kucsko2013nanometre}. 

When exposed to an external magnetic field, the degeneracy of the $m= \pm 1$ states of the NV-centre is lifted and Zeeman splitting can be measured optically.
This can be used to sensitively detect magnetic fields~\cite{maze2008nanoscale}. Single NV-centers have been used to detect nanotesla magnetic fields at nanoscale~\cite{staudacher2013nuclear}, while ensembles of NV-centres have allowed the detection of picotesla fields~\cite{wolf2015subpicotesla,bennett2021precision}. 
This sensitivity to local magnetic fields has been applied to measure the spin density of biomolecules and reconstruct three dimensional structures~\cite{balasubramanian2008nanoscale}, as well as to detect the action potential of a single neuron in a living worm~\cite{barry2016optical} (see figure \ref{fig:NV}b). Similar action potential measurements have also been made using atomic-ensemble-based quantum magnetometers, which also function by monitoring Zeeman splitting~\cite{jensen2016non}. 

%Quantum photon probes
In the previous few paragraphs we have discussed how the use of quantum effects in nanoprobes can allow new and improved biosensing technologies. Quantum effects in the optical fields used to measure biological specimens can also be used to improve performance~\cite{taylor2016quantum}. 
Light is widely used in biosensors, such as all forms of optical microscopy, interferometric biosensors~\cite{mauranyapin2017evanescent,piliarik2014direct,dantham2013label}, plasmonic devices~\cite{pang2012optical,zijlstra2012optical}, etc... The performance of such sensors generally improves with increasing light intensities, allowing both stronger interactions with the specimen and lower noise.
However, biological samples are  known to be sensitive to damage introduced by light.
Various forms of damage exist,  including photochemical effects, local heating, and physical damage, and these can affect the growth, viability and function of the specimen~\cite{sowa2005direct, landry2009characterization, waldchen2015light}.

By constraining the optical intensities that can be used in biosensing, photodamage places limits on performance that can only be overcome by exploiting quantum correlations between photons~\cite{taylor2016quantum}. At a fixed light intensity, the sensitivity of an optical biosensor that uses classical light -- by which we mean light that does not exhibit quantum correlations  --
is fundamentally limited by quantisation of light, or {\it shot noise}~\cite{mauranyapin2017evanescent,mauranyapin2019quantum,piliarik2014direct}. For instance, it can be immediately seen that the photon nature of light will constrain the size of molecule that could be detected via light scattering: if the molecule is small enough that it does not scatter at least one photon within the measurement period, then in the absence of quantum correlations it is fundamentally undetectable. It is somewhat remarkable that this statement is not more generally true -- that is, that quantum correlations {\it do} allow the detection of a molecule that, on average, scatters less than one photon. Nevertheless, this is indeed the case~\cite{alexander2020sounds}.  
One way to appreciate this is to consider the wave, rather than particle, picture of light. Detecting the wave-like properties of the scattered light (as can be done using a homodyne receiver) rather than discrete photon arrival events, we observe a continuous electric field with amplitude that is finite even if less than one photon is on-average  scattered. 
Using quantum correlated light can reduce the noise on this electric field. With sufficient noise reduction very small optical electric fields can be detected -- even if the energy in the electromagnetic field is much less than the energy of a photon.

Without quantum correlations, the sensitivity of an optical biosensor can be improved in a variety of ways. More intense light can be used, the molecule could be placed within a cavity, or a contrast-enhancing label could be attached to the molecule~\cite{vollmer2008whispering,baaske2014single,baaske2016optical,kim2017label}. In each case, this works by increasing the level of optical scattering from the molecule.
However, it is not always possible or desirable to label biomolecules, since this can change their behaviour or biochemical environment~\cite{szabo2018effect,sanchez2017effects,hughes2014choose}. Nor is it often feasible to place the biomolecule within a cavity, or to use arbitrarily high optical intensities. Thus, quantum correlations can play an important role, allowing improved performance without intruding on the specimen through increased intensity or the presence of a label.

Quantum correlations have been applied to reduce measurement noise in a range of biosensing experiments~\cite{taylor2016quantum}. 
This allows improved signal-to-noise at fixed optical illumination intensity, and therefore improved performance without increased optical intrusion on the specimen.
Here we focus on a specific class of quantum correlated light, termed {\it squeezed light}, which has been applied in several biological experiments and tends to 
outperform other non-classical states in biotechnological applications~\cite{taylor2013biological, taylor2016quantum}.
A range of processes exist to engineer squeezed states of light, the most popular being use of an optical parametric oscillator ~\cite{andersen201630}. These processes introduce quantum correlations that can reduce the noise on  the optical amplitude or phase quadrature of the light~\cite{taylor2016quantum}. Phase noise variance reductions beneath the shot noise as large as a factor of thirty ($-15$~dB) have been achieved with squeezed light~\cite{vahlbruch2016detection}, and have been successfully applied in detection of gravitational waves with kilometre-length interferometers~\cite{aasi2013enhanced}.

\begin{figure}[!ht]
 \centering
 \includegraphics[width=1\linewidth]{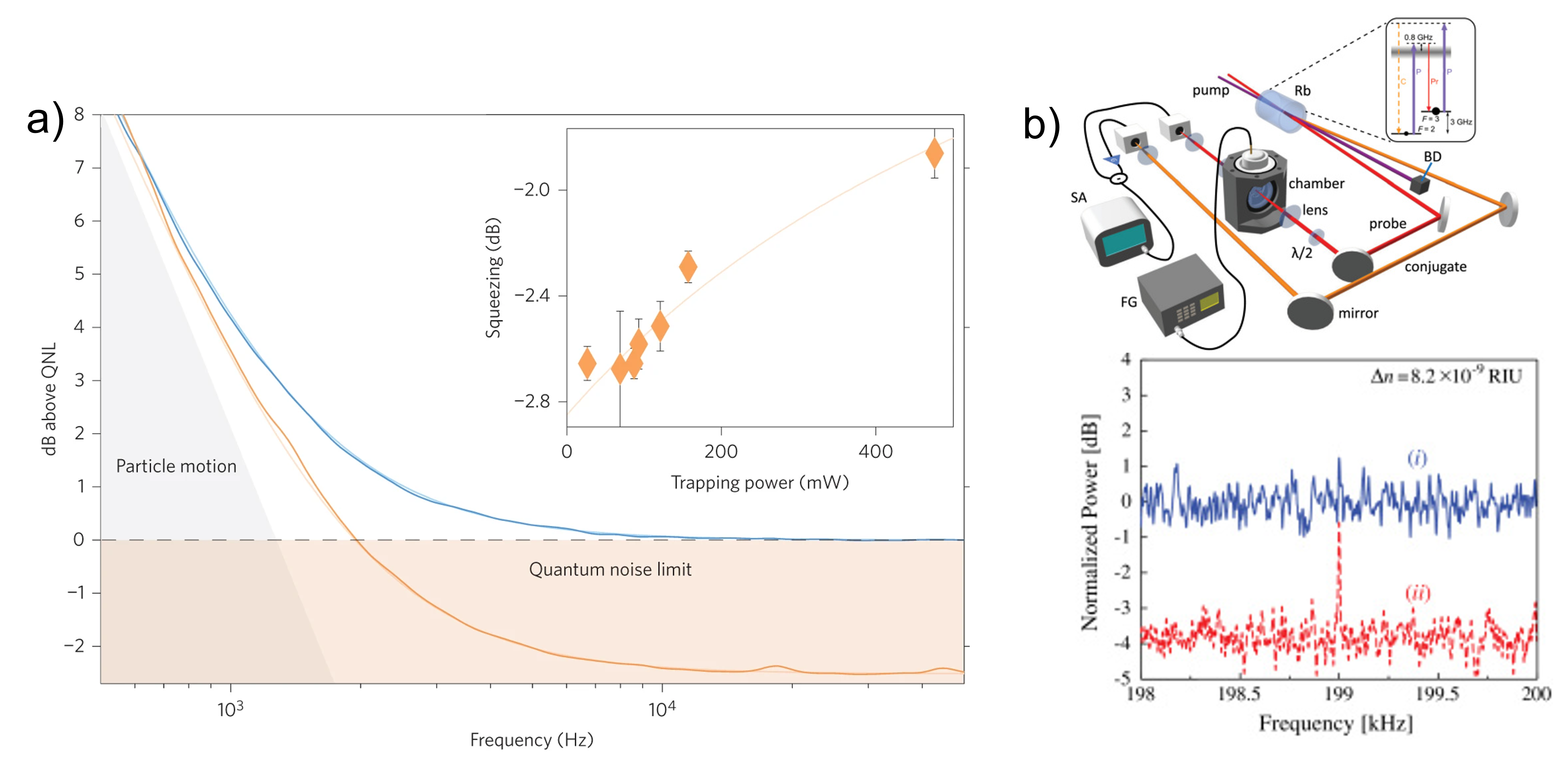}
 \caption{Biosensing with quantum squeezed light. a) SNR enhancement of particle tracking in optical tweezers. Measured mechanical noise spectra of a 2~$\mu$m trapped silica particle with classical (blue) and with squeezed (orange) light. This shows the noise reduction achievable with squeezed light. Inset, amount of squeezing as function of optical trapping power. (reused from Ref.~\cite{taylor2013biological} with permission). b) SNR enhancement in plasmonic sensor particle detection. Top optical setup used to create amplitude squeezed light that passes trough the sample chamber. Bottom, measured spectrum with classical light (blue) and squeezed light (red). The signal, as a modulation at 199~kHz, can only be detected with quantum light. (Reused with permission from Ref.~\cite{dowran2018quantum}. Copyright (2018) The Optical Society)} 
\label{fig:squeezed}
\end{figure}

In biotechnology, quantum squeezed states have been used to improve particle tracking inside yeast cells~\cite{taylor2013biological} (see figure  \ref{fig:squeezed}a), to enhance the spatial resolution of photonic force microscopes~\cite{taylor2014subdiffraction}, and to enhance laser beam positioning for atomic force microscopy~\cite{treps2003quantum, treps2004nano, pooser2015ultrasensitive, pooser2020truncated}. In the first of these examples, the improved particle tracking allowed more precise measurements of the viscoelasticity of local regions within the cell. Accurate measurements of viscoelasticity are  important for biotechnological applications since changes in viscoelasticity can be used to discriminate cancerous and healthy cells~\cite{nematbakhsh2017correlating}; in the development of pharmaceuticals, food and cosmetics~\cite{gallegos1999rheology}; and more fundamentally, can be linked to complex active reorganization of the cytoplasm, as well as protein folding and movement~\cite{madsen2021ultrafast}.
Squeezed states have also been applied to optical plasmonic biosensors resulting in a 56\% sensitivity enhancement of nanoparticle detection~\cite{dowran2018quantum,fan2015quantum} (see figure \ref{fig:squeezed}b). This opens a path toward the detection of analytes faster and at lower concentrations than is otherwise possible.

As well as reducing noise, quantum correlations can also be used to increase the signal generated by the sample of interest. For example, in the case of optical phase measurement via the interference of two beams, when exactly $N$ photons are used  in an entangled \textit{NOON state}, 
the phase shift created by the sample is increased by a factor $N$~\cite{taylor2016quantum}. This effectively results in sub-wavelength interference fringes impossible to achieve with classical light. Proof-of-principle biosensing experiments have been performed with NOON states to detect protein concentration~\cite{crespi2012measuring} and sucrose hydrolysis~\cite{peng2021real}. However, it is challenging to both create NOON states with more than five photons and  produce a sufficiently high flux of NOON states to achieve absolute performance advantages~\cite{taylor2016quantum}. Moreover, NOON states are quickly degraded by losses during optical propagation throughout the sensor. 
Practical biosensing applications of low-flux pairs of quantum correlated photons have nevertheless been demonstrated. For example, in Ref.~\cite{phan2014interaction} pairs of entangled photons are used to study a single retina rod cell. In this work, one photon is sent to the rod cell and its arrival is heralded by the detection of the second photon. This reduces the statistical uncertainty on the photon arrival on the rod cell. The single photon response of the rod cell and it transient response can therefore be monitored without having to resort to statistical modeling that would be needed when a classical source of light is used.

%Outlook
We have seen that biosensing can benefit from quantum technology in different ways: by exploiting quantum processes, by introducing  quantum probes and by exploiting quantum correlations. While several quantum biosensing techniques have been employed in biotechnology for decades -- most particularly NMR -- many new techniques and approaches are being demonstrated, and beginning to transform what it is possible to sense in biological systems~\cite{xavier2021quantum}. One particular application of these approaches is in bioimaging, as discussed in the following Section.
We anticipate other applications ranging from real-time measurements on single proteins that can image their structure and resolve their folding, conformation changes, chemical reactions and chirality~\cite{xavier2021quantum}, to quantum bio-science laboratories on a chip which could increase our knowledge of quantum processes in biology~\cite{xavier2021quantum} (see Section \ref{sec:QuantumEffectsBiology}).

\section{Quantum-enabled bioimaging}

Precision imaging tools are a key enabler of progress in biotechnology.
They are used routinely for medical diagnosis of vascular diseases~\cite{lao2008noninvasive,taruttis2015advances, perekatova2021quantification} and brain conditions \cite{pfefferbaum1995longitudinal,politi2020magnetic, debette2019clinical}, and for monitoring of living tissues \cite{chan2011basics, shang2017development}.  By allowing  the structure and activity of neural networks to be studied and controlled both {\it in vitro} and {\it in vivo} \cite{adam2019voltage,piatkevich2019population,yang2016subcellular}, they open a crucial door to understanding of brain function and brain diseases, while providing similar capability for studies of complex cancers \cite{han2020cancer}. They allow drug delivery to be monitored in real time \cite{saar2011imaging, dolovich2004imaging, vanden2019raman} and, as such, are crucial for the development of pharmaceuticals~\cite{willmann2008molecular}.
Drug development is an important example of an application of imaging in biotechnology. It is a long and risky process: only 1 out of 10,000 proposed compounds reach the market in the United States~\cite{zambrowicz2003knockouts} after an average of more than 14 years of research and development~\cite{dimasi2001new} at a punishing cost of up to 2.8 billion USD \cite{dimasi2003price,wouters2020estimated}.  {\it Molecular imaging}, a discipline that characterizes molecular processes in living organisms using complementary imaging techniques, has been used to speed up pre-clinical and clinical processes in drug development, allowing reductions in cost and time, while providing higher levels of proof of drug efficacy~\cite{willmann2008molecular,huang2019molecular}.

Existing imaging systems face both fundamental and technological limitations~(e.g. \cite{taylor2013biological}) which constrain the breadth and quality of their biotechnological applications. For instance, due to limitations in resolution and sensitivity, much remains unknown about how molecules are transported and distributed within and between cells~\cite{alejandro2020manganese, van2018subcellular,nazari2020transport}, and how they regulate cellular-scale processes~\cite{perreault2018intracellular,fujita2007regulation}. 
Similarly, our understanding of the brain is limited by the resolution of neuroimaging technique such as Magnetoencephalography (MEG) \cite{cohen1972magnetoencephalography}. MEG detects currents in neural circuits via the magnetic fields they produce. However, it generally requires cryogenic temperatures and has resolution orders of magnitude above the size of a neuron~\cite{hamalainen1993magnetoencephalography,iivanainen2019scalp}. How brain function arises from collections of interconnected neurons is perhaps the most significant question in neuroscience~\cite{bassett2017network}. How neurodegenerative disorders such as Alzheimer and Parkinson progress is an equally important question in biomedicine~\cite{tan2021interdisciplinary,sado2018estimated}, as is understanding the effects of pharmaceuticals on neural circuits~\cite{castren2017neuronal,browning2019electroshock}. Limitations in our ability to observe action potentials at neuron-level and in real time, are a major barrier to answering this questions. 
It is similarly important to improve the sensitivity and resolution of optical imaging techniques \cite{luker2008optical}, such as bioluminescence, fluorescence, photoacoustic imaging \cite{contag2002advances, alsawaftah2020bioluminescence,woll2017super,attia2019review}. Here, as already discussed in the previous section in the context of quantum sensors, photon budget is a major limitation, with optical irradiation influencing the environment, growth and function of the sample~\cite{waldchen2015light,fu2006characterization,schermelleh2019super}, and ultimately its viability~\cite{waldchen2015light,schermelleh2019super}.
For example, Raman microscopy is a technique that selectively images molecular bonds and has key application in biotechnology such as monitoring antibiotic response \cite{schiessl2019phenazine} or nerve degeneration \cite{tian2016monitoring}. State of the art Raman microscopes are already limited in speed and resolution by photodamage \cite{camp2015chemically,fu2006characterization}, constraining future applications of the technique in biotechnology \cite{chrimes2013microfluidics,palonpon2013molecular}.

Quantum imaging technologies can both improved in existing capabilities, and open the door to measurements and applications that would otherwise not be possible. Unique quantum effects associated with nanoscale objects \cite{degen2017quantum} can be utilised to increase measurement sensitivity \cite{broadway2018quantum}, extend the resolution of imaging techniques to new scales \cite{golman2003molecular,chen2018technique}, or enable existing technology to operate a room temperature while preserving their performance, increasing their ability to be used at point of care. Quantum correlations and entanglement can also be used to resolve smaller features and increase performance~\cite{sewell2013certified}, for example improving  optical imaging at a fixed photon budget \cite{Casacio2021}. We re-emphasize here that two imaging techniques based on quantum effects are routinely used: Magnetic resonance imaging (MRI) and Positron emission tomography (PET) \cite{kuhl1963image,phelps2000positron}. In the remainder of this section we provide some examples of key newly developed and developing quantum imaging technologies.

Magnetic Resonance Imaging (MRI) \cite{edelman2014history} has became a key technique for medical imaging of soft body tissues and organs. MRI functions in a similar way to NMR, described in the previous section, monitoring the response of nuclear spins in a subject to high-frequency radio pulses in the presence of a strong magnetic field.
Contrast agents are used to improve the sensitivity and resolution of conventional MRI. These agents raise toxicity concerns \cite{jung2015molecular}, and even when using them the resolution of MRI is limited to around tens of micrometers \cite{boretti2019nitrogen}, unsuited for molecular scale imaging \cite{glover2002limits}. 

\begin{figure}[!ht]
 \centering
 \includegraphics[width=0.85\linewidth]{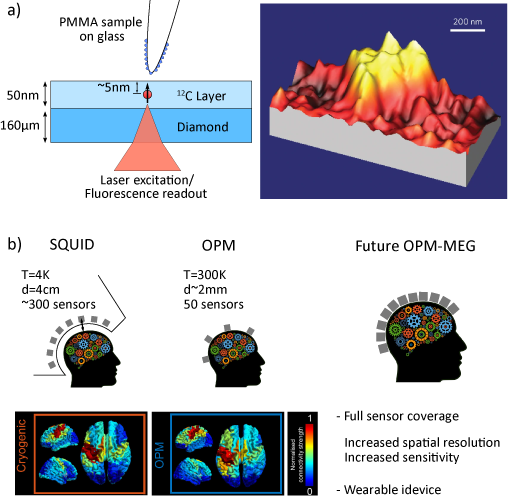}
 \caption{Quantum magnetic bioimaging a) Nano-MRI. Left: setup use generate the nano-MRI image of a single proton. A PMMA polymer sample attached to fork is brought into contact with substrate containing a near-surface NV centre. The protons in the sample are detected by the NV centre as the sample is scanned past it. The magnetic state of the NV centre is read out optically via spin-dependent fluorescence. Right: PMMA samples image with a single NV centre. The height shown is proportional to the NMR signal s(x,y). Apparent roughness is primarily due to photon shot noise. (right image reused from Ref.~\cite{rugar2015proton} with permission). b) Current MEG technologies and future promise of optically pumped magnetometers (OPM). Left: schematic of current superconducting quantum interference device (SQUID)-MEG and OPM-MEG systems together. Right: future OPM-MEG systems may include a high number of densely packed sensors. Such systems could provide better signal-to-noise ratio (SNR) and spatial resolution, while adding the capability to conduct experiments in more naturalistic settings (Scan images reused from Ref.~\cite{boto2021measuring}, license CC BY)} 
\label{fig:magnetic_bioimaging}
\end{figure}

Nanoscale MRI is a set of recent techniques that seek to achieve nanoscale resolution, and open a path to molecular biology applications~\cite{boretti2015towards}.
Such techniques require nanoscale magnetic probes.
However, most of these probes are only able to operate at cryogenic temperatures due to their short dephasing times under ambient conditions ~\cite{boretti2015towards}. 
This limitation can be overcome using the quantum defects in diamond (typically NV centres) described in the previous section, paving the way towards nano-MRI in living organisms~\cite{boretti2015towards,degen2008scanning}.
Quantum diamond-based nano-MRI has recently been applied to two dimensional imaging of the magnetic field emanating from a single proton~\cite{rugar2015proton} with a reported resolution of 12~nm see figure \ref{fig:magnetic_bioimaging}a. 
Here, a single NV center near the surface of the diamond was made sensitive to the proton magnetic field by using microwave pulses to manipulate the NV spin state \cite{rugar2015proton,staudacher2013nuclear}. Close proximity to the proton results in a dip in the NV spin coherence, inversely proportional to the NV-nuclei distance, which allowed the construction of an image by raster scanning. This represents a first step toward applications to bioimaging, with several developments proposed to improve the signal-to-noise and extend the demonstration to biological samples \cite{siyushev2010monolithic,babinec2010diamond,mamin2014multipulse,mamin2012high}.

Biomagnetic imaging is useful to many key applications in biotechnology  \cite{morris1986nuclear,taylor2016quantum}. However, available techniques either lack the spatial resolution required for subcellular measurements, or aren't compatible with living systems \cite{le2013optical}. Recent advances in magnetic imaging have seen these gaps filled though the development of  wide-field quantum diamond microscopes \cite{le2013optical}.
Quantum diamond microscopes combine an optical microscope with a diamond chip that contains 
a thin layer of Nitrogen-Vacancy (NV) centers  and is placed beneath the sample~\cite{healey2020comparison}. They can be diffraction limited, and because they probe the magnetic field outside of the sample they are intrinsically biocompatible~ \cite{le2013optical}.
The function of quantum diamond microscopes relies on the ability to using light and microwaves to control the states of NV centres at room temperature and to 
optically read them out, as described in the previous section~\cite{gruber1997scanning}.
This has allowed the first wide-field magnetic images of living cells with sub-micrometer resolution~\cite{le2013optical, glenn2015single,mizuno2018wide,foy2020wide}. It has also allowed dynamical tracking of near-DC fields and high frequency magnetic fields  \cite{levine2019principles}. Moreover, the sensitivity of the NV-center to temperature enables simultaneous magnetic and thermometric imaging \cite{foy2020wide}. Quantum diamond microscopes have been used to resolve cancer biomarkers corresponding to rare tumor cells within a large population of healthy cells \cite{glenn2015single}, and have been combined with MRI \cite{davis2018mapping} to bridge the gap between macroscopic and microscopic effects in bioimaging.

Functional neuroimaging of human brain activity is a staple of medical imaging and neuroscience. Magnetoencephalography (MEG) \cite{cohen1968magnetoencephalography} is a powerful and non-invasive tool that detects the extracranial magnetic field generated by neuronal activity in our brain. These femtotesla-scale field measurements have been made possible by superconducting quantum interference devices (SQUIDs)~\cite{cohen1972magnetoencephalography} compatible with the weak signals from electrical currents in the brain. This provides real-time images of brain activity, with wide applications in medical diagnosis and neuroscience \cite{baillet2017magnetoencephalography, stefan2017magnetoencephalography, dunkley2015low, lopez2018role}. 
However, the  need to cryogenically cool SQUIDs limits MEG tools to specialised facilities \cite{boto2021measuring}, increases the distance  between the subject and the sensor to around 2~cm \cite{hill2020multi}, makes small movements of the subject problematic \cite{gross2013good}, and overall increases the cost of MEG systems \cite{boto2021measuring}.
Recent progress in atomic magnetometry is addressing these limitations\cite{shah2013compact}.
Atomic magnetometers rely on the manipulation and readout of atomic spins. A high pressure gas is laser-pumped into a magnetically sensitive state \cite{tierney2019optically}. Much like a quantum diamond magnetometer, the presence of a magnetic field Zeeman-shifts the atomic energy levels.
These shifts can be precisely read-out, for example by a polarisation measurement on a probe beam \cite{colombo2016four}. Recent advances in microfabrication \cite{griffith2010femtotesla, shah2013compact} and improvements in sensitivity have enabled atomic magnetometers to reach and even surpass the sensitivity of SQUIDs \cite{budker2000sensitive, bennett2021precision}, and opened up applications and in brain imaging \cite{tierney2019optically}. Atomic magnetometers work at room temperature and be can miniaturised to sizes considerably smaller than SQUID magnetometers. They offer increased spatial resolution compared to SQUIDs~\cite{iivanainen2019scalp}, can be adapted to different head size and account for head movement \cite{hill2019tool}, and can potentially reduce costs \cite{boto2021measuring}. Optically-pumped magnetometers, a type of atomic magnetometer-based sensors, are already commercially available in a wearable form \cite{hill2020multi,boto2021measuring} (e.g. from QuSpin) and 
 have already enabled new neuroscience studies \cite{roberts2019towards, hill2019tool,boto2021measuring}. Other types of quantum magnetometers compatible with MEG are under development, such as NV-center based sensors  \cite{schneiderman2019scalp} and optomechanical sensors \cite{li2021cavity, li2018quantum}. Such advances will permit further reductions in  weight, size and energy consumption, and potentially make MEG available in the doctor's office (see figure \ref{fig:magnetic_bioimaging}b). 

Optical microscopy and more specifically fluorescence microscopy is a crucial tool used to investigate cells, living tissues and small organisms \cite{li2017aggregation, zhao2017recent, etrych2016fluorescence}. While it is compatible with live-cell imaging, the optical diffraction limit of simple imaging tools prevents the structure of objects that are smaller than around the optical wavelength from being resolved~\cite{abbe1873contributions}: two light sources such as fluorophores that are separated by less than a few hundred nanometres will be indistinguishable. Most biological structures in cells are smaller than this, which limits many applications of optical imaging in biotechnology. 
Various super-resolution techniques have been developed to overcome the diffraction limit \cite{hell1994breaking,zipfel2003nonlinear,gustafsson2000surpassing, rust2006sub, betzig2006imaging,dertinger2009fast}. 
Stochastic Optical Reconstruction Microscopy (STORM) is a fluorescence technique that achieves resolution that is around a factor of ten beyond the diffraction limit \cite{rust2006sub} by consecutively activating small subsets of fluorophores so that at any given time the emission from each activated fluorophores is spatially separated. This allows the position of each flurophore to be determined with accuracy beyond the diffraction limit
 using centroid estimation. Repeating the process on different subsets of the fluorophores, once combined, provides a high-resolution image (see for example figure \ref{fig:optical_bioimaging}a). STORM is an active field of research, limited in practice by the properties of the fluorescent probe  and the required acquisition time \cite{samanta2019organic}. While STORM has broad applications in bioimaging, these limitations ultimately constrain its resolution and its application \cite{samanta2019organic}. 
 Quantum effects can be used to extend the resolution of STORM resolution as well as other super-resolution techniques.
For instance, the anti-bunching properties of the photons emitted by fluorophores \cite{kimble1977photon}  have been demonstrated to enhance the resolution and contrast of imaging \cite{israel2017quantum, assmann2018quantum}. This relies on using the photons statistics to identify how many emitters exist at a given pixel and requires only relatively minor setup modifications: fast single-photon resolving avalanche photodiode arrays \cite{vitali2014single} or fiber bundle cameras \cite{israel2017quantum} instead of CCDs~\cite{assmann2018quantum}. Successful Quantum-enhanced STORM has been demonstrated \cite{assmann2018quantum}, along with other super resolution techniques \cite{tenne2019super}, paving the way to potential real-time molecular imaging.

As discussed in the previous section, quantum correlations between photons can also be used to reduce the noise on optical measurements, and therefore improve sensitivity at a fixed photon budget. This is particularly important for optical bioimaging technologies, where photodamage is a prevalent problem~\cite{waldchen2015light,fu2006characterization,schermelleh2019super}.
Recently, it has been demonstrated that quantum squeezed light (see previous section) can allow contrast in bioimaging that is fundamentally beyond the possibilities of classical techniques~\cite{Casacio2021} -- that is, the biological specimen would be exposed to damage at intensities below those required to achieve the contrast of the quantum image. This is an example of absolute quantum advantage~\cite{arute2019quantum}, offering the ability to see biological structures that would otherwise be fundamentally beyond reach. This result was achieved using a Raman microscope, a widely used bioimaging technique  \cite{cheng2015vibrational,camp2015chemically} that is highly specific \cite{wei2017super} and doesn't require markers, enabling the study of biological processes such as metabolic processes \cite{zhang2019forster} nerve degeneration \cite{tian2016monitoring} or antibiotic responses \cite{schiessl2019phenazine}. The contrast and speed of state-of-the-art Raman microscopes is already limited  by photodamage \cite{fu2006characterization,camp2015chemically}, so that quantum enhanced performance is extremely relevant. In the work of Ref.~\cite{Casacio2021}, the use of squeezed illumination light enabled a 35\% enhancement of signal-to-noise ratio, or contrast, and applied this to image living cells \cite{Casacio2021}. It is to be noted that non-biological squeezed light reach factors up to 30 in signal-to-noise enhancement \cite{vahlbruch2016detection}, the same squeezing techniques could be used to greatly improve the performance of \cite{Casacio2021}. 
This result opens the door for further applications of Raman microscopy and optical imaging in general such as video-rate imaging of weak molecular vibrations and
label-free spectrally resolved imaging \cite{cheng2015vibrational,wei2017super}.

\begin{figure}[!ht]
 \centering
 \includegraphics[width=0.85\linewidth]{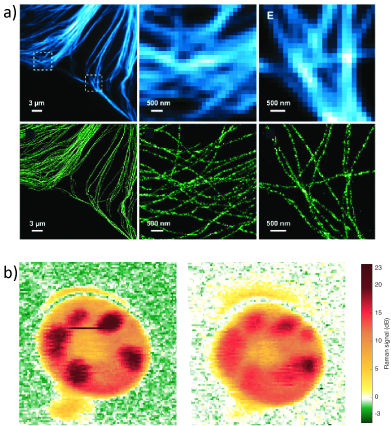}
 \caption{Quantum optical bioimaging a) STORM imaging of microtubules in a living cell. Top left: conventional immunofluorescence image of microtubules in a large area of a mammalian cell. Bottom left: STORM image of the same area. (Top center and right) conventional and (bottom center and right) STORM images corresponding to the boxed regions on top left. (reused from Ref.~\cite{bates2007multicolor} with permission, license 5177370549654). b) Comparison of a shot noise limited (right) and quantum enhanced (left) Stimulated Raman images of a live yeast cell (Saccharomyces cerevisiae) in aqueous buffer at 2,850~cm$^{-1}$ Raman shift. Several organelles are clearly visible. The faint outline of what may be the cell membrane or wall is also visible, showing that the microscope has a resolution of around 200 nm. The measurement noise of the quantum-enhanced image is reduced by 1.3 dB below shot noise, corresponding to a 35 percent signal-to-noise ratio and contrast improvement (Reused with permission from Ref.~\cite{Casacio2021}. }
\end{figure}
\label{fig:optical_bioimaging}

In this section we have seen that quantum technologies are able to extend the performance of existing biological imaging techniques, and allow imaging applications that would otherwise be inaccessible. 
Indeed, the quantum imaging technologies of MRI and PET have been in use in biological imaging for decades \cite{kuhl1963image,phelps2000positron}.  Recent technological advances such as biocompatible quantum nanoprobes \cite{schirhagl2014nitrogen}, room-temperature quantum  magnetometers~\cite{boto2017new}, and optical microscopes with quantum-enhanced performance~\cite{Casacio2021} extend beyond these important early applications. 
We anticipate these advances, and further progress that stem from them, will over time prove to be transformative in a wide variety of areas of biotechnology.
 For example, quantum-enabled real-time, long-duration images of action potential dynamics in our brain could provide more detailed understanding of neuronal plasticity and even of 
how consciousness arises from neurons \cite{crick2003framework}, while room-temperature quantum-MEG could transform how brain diseases are treated~\cite{schneiderman2019scalp,wald2020low}. In much the same way, we anticipate that improved nanoscale imaging of the dynamics of single-molecules will provide important insights into protein-protein interactions, protein folding, and the efficacy of drugs and drug delivery, among other areas
 \cite{willmann2008molecular,huang2019molecular,boretti2015towards}.

\section{Quantum control of biomolecular dynamics}

\begin{figure}[!ht]
 \centering
 \includegraphics[width=1\linewidth]{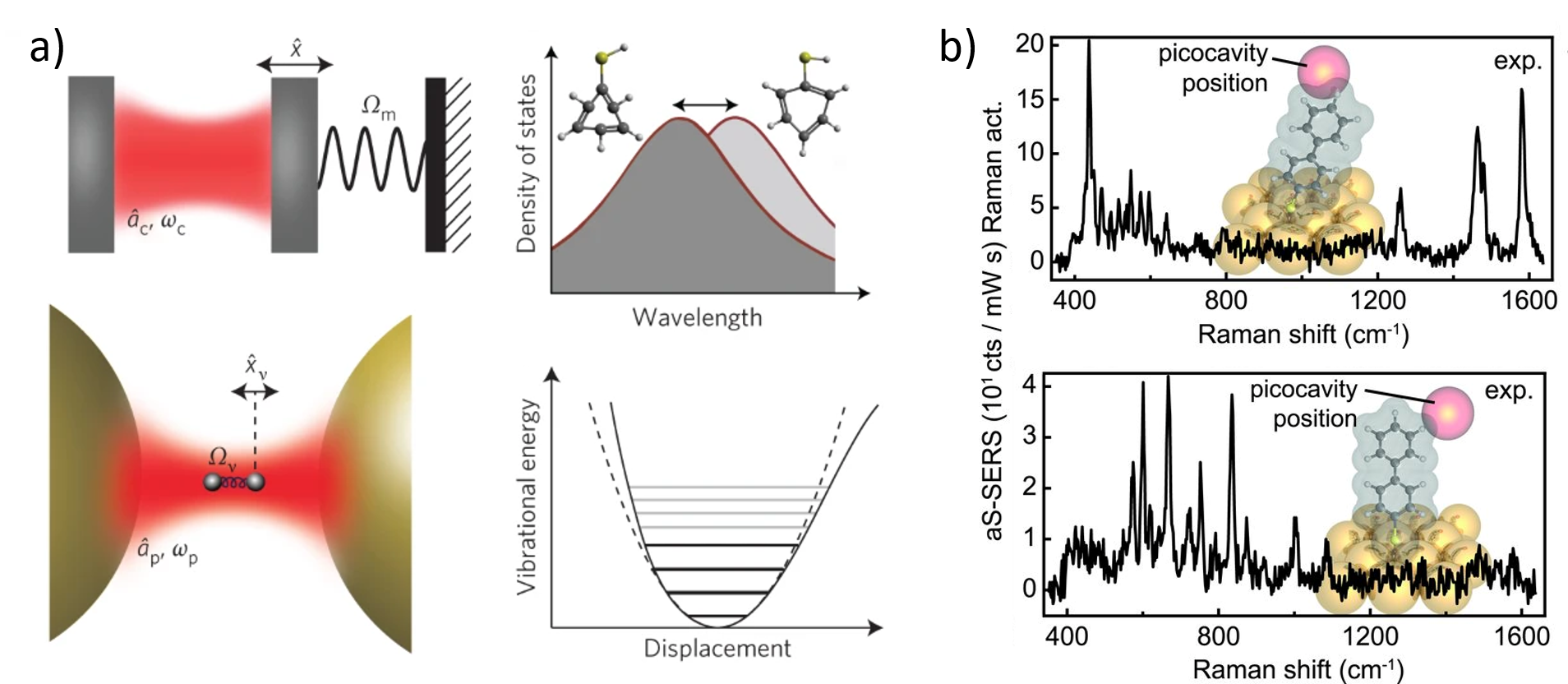}
 \caption{Molecular optomechanics. a) Cavity optomechanical model of the interaction between a molecule inside a plasmonic cavity. Top left, typical optomechanics model containing a mechanical oscillator and an optical cavity. Bottom left, optical field in the plasmonic cavity.Top right, schematic of frequency shift due to mechanical vibration.  Bottom right, vibrational energy level of the mechanical oscillator.  (Reproduced from Ref.~\cite{roelli2016molecular} with permission 5176190084626). b) Plasmonic enhanced Raman signal. Measured Raman spectra of  biphenyl-4-thiol molecule sandwiched between a gold film and a gold nanoparticle. Top and bottom are spectra for two different positions of the nanoparticle.  (Reproduced from Ref.~\cite{benz2016single} with permission 5176190385790).} 
\label{fig:molecularOptomechanics}
\end{figure}

Many biotechnological applications rely on the ability to control chemical reactions to obtain a desired product with high yield. Usually, this is achieved by varying bulk parameters such as the temperature, pressure and acidity, or by introducing catalysts. However, in 1995 it was shown that appropriately tailored optical pulses could allow coherent control, by driving coherences between different electron excitation pathways~\cite{zhu1995coherent}. This began the field of coherent control. Coherent control offers the possibility to not only improve the yield of chemical reactions, but also drive chemical reactions that would otherwise not occur, creating entirely new chemical species~(see e.g. \cite{keefer2018pathways,shapiro2012quantum}). It is now a mature field. Potential applications both in understanding chemical reactions and in biotechnology have provided the impetus for rapid experimental progress, with demonstrations of coherent control of photoexcitation~\cite{zhu1995coherent} and photodissociation~\cite{assion1998control}, of the energy transport in photosynthetic light harvesting~\cite{herek2002quantum}, of isomeriation of retinal cells~\cite{prokhorenko2006coherent}, of chemical bonding~\cite{levin2015coherent} and of photochemical reactions~\cite{blasing2018observation}, among other processes. Several excellent reviews of coherent control
already exist (e.g. \cite{keefer2018pathways, silberberg2009quantum, hishikawa2020ultrafast}), including 
the comprehensive textbook, Ref.~\cite{shapiro2012quantum}. However, ideas from the field have inspired other newer approaches to the biomolecular control, particularly in the area of  {\it molecular polaritonics} which we focus on here~\cite{herrera2020molecular,sanvitto2016road}.

In the field of molecular polaritonics laser driving is combined with a nanoscale optical or plasmonic resonator (see Fig.~\ref{fig:molecularOptomechanics} a)). The  resonance can be used to modify the molecular density of states, changing the states that molecules are able to transition in to, and thereby providing a new control handle. The intensity build-up in the resonator, combined with nanoscale confinement of light, offers the prospect of quantum control at the single molecule level, evading the heterogeneity and ensemble averaging that often obscures underlying single-molecule effects. Furthermore, the time delay associated with the build-up of intensity introduces a time lag in the response to molecular dynamics. This provides the possibility of dynamical backaction between field and molecule~\cite{bowen2015quantum}, which could be used to cool molecular degrees of freedom, or to control them in a variety of other ways.

Molecular polaritonics developed from cavity quantum electrodynamics, where the electronic state of an atom within a small optical cavity is hybridised with the state of the field in the cavity~\cite{walther2006cavity,aoki2006observation}. Here, the atom and field interact via an electric dipole interaction. Due to quantum zero-point energy, this interaction occurs even when there are no photons in the cavity. The interaction rate increases as the volume of the cavity is reduced. Once the vacuum interaction rate (or vacuum Rabi frequency) exceeds the dissipation rates of the cavity and atomic  transition, the system reaches a {\it strong coupling} regime where the energy eigenstates hybridise into polaritonic states that combine the properties of light and atom (and electron if a plasmonic resonator is employed). When applied to a molecule, rather than an atom, this fundamental change to the energy landscape can have a large influence on the molecular dynamics~\cite{herrera2020molecular}.

Much work in molecular polaritonics has focused on  electronic degrees of freedom of ensembles of molecules. The use of an ensemble increases the interaction strength, allowing the use of larger optical cavities and providing more straightforward access to the strong coupling regime~\cite{torma2014strong, garcia2021manipulating}. 
It has been shown that it is possible to employ strong coupling to vacuum fields to modify chemical energy landscapes and chemical reactions~\cite{hutchison2012modifying}, and that the conductivity of organic polymers can be greatly enhanced when hybrised with a vacuum electromagnetic field~\cite{orgiu2015conductivity}. Strong coupling has been switched on and off by photochemically inducing molecular conformational changes~\cite{schwartz2011reversible}. Room temperature strong coupling between a single light-emitting dye molecule  and the cavity mode of a plasmonic nanoparticle has been achieved~\cite{chikkaraddy2016single}, as have 
room temperature polaritonic condensates that exhibit superfluidity ~\cite{lerario2017room}, 
among much other progress.

Recently, the new area of {\it molecular optomechanics} has developed, combining the fields of molecular polaritonics and quantum optomechanics~\cite{bowen2015quantum}. Quantum optomechanics studies the radiation-pressure interaction between light and a mechanical resonator
(Fig.~\ref{fig:molecularOptomechanics} a))%
~\cite{bowen2015quantum}.  Ultraprecise optomechanical sensors have enabled the detection of gravitational waves in kilometer-scale interferometers~\cite{abbott2016gw151226}, as well as on-chip sensing of magnetic fields~\cite{forstner2012cavity, yu2016optomechanical, li2020ultrabroadband,li2018invited}, acceleration~\cite{krause2012high, zhou2021broadband, huang2020chip}, forces~\cite{gavartin2012hybrid, harris2013minimum}, temperature~\cite{purdy2017quantum} and  ultrasound~\cite{basiri2019precision,westerveld2021sensitive}; while quantum optomechanical control techniques have enabled cooling of
mechanical vibrations~\cite{metzger2004cavity, harris2016laser, lee2010cooling, naik2006cooling, schafermeier2016quantum}, even to the quantum ground state~\cite{chan2011laser, delic2020cooling}, amplification and lasing of mechanical motion~\cite{carmon2005temporal, lin2009mechanical, taylor2012cavity, bekker2017injection}, strong-coupling between light and mechanical degrees of freedom~
\cite{groblacher2009observation},
and engineering of exotic vibrational quantum states~\cite{satzinger2018quantum, riedinger2016non}.

The vibrational motion of a biomolecule within the field of a plasmonic resonator or optical cavity can play an equivalent role to the mechanical resonator in a canonical quantum optomechanical
system~\cite{benz2016single,roelli2016molecular} (Fig.~\ref{fig:molecularOptomechanics} a)). Here, rather than electronic states as we discussed earlier,  molecular vibrations are hybridised with the optical cavity field. This occurs via the Raman interaction, where a vibrational mode of the molecule couples two optical frequencies that have a frequency difference equal to the vibrational frequency. Through this mechanism it is in principle possible to  control the vibrational state of the biomolecule in the same manner as a more macroscopic mechanical resonator. This opens up the possibility to apply the quantum measurement and control techniques developed for quantum optomechanics into molecular biotechnology. 
Following this roadmap, room temperature strong coupling of the ground state vibrational levels of an ensemble of molecules to a cavity was demonstrated in 2015 \cite{shalabney2015coherent, george2016multiple}.
Vacuum strong coupling has since been shown to have a  significant influence on chemical reactions~\cite{thomas2016ground}, controlling the reactivity landscape~\cite{thomas2019tilting} and allowing reaction rates to be both promoted and suppressed. Strong coupling has also allowed quantum interference of vibrational pathways~\cite{muller2018nanoimaging}, and has reached the single molecule level (Fig.~\ref{fig:molecularOptomechanics} b))~\cite{benz2016single}.

Compared to the direct dipole coupling of molecular polaritonics with electronic degree of freedom,  the presence of two optical frequencies when coupling to molecular vibrations offers additional  flexibility. It allows specific molecular resonances to be targeted through judicious choice of detuning between the cavity resonance frequency and an optical drive tone, allows the properties of hybrid modes to be tuned by using light to pump one of the optical modes, and increases the 
range of processes that can be implemented~\cite{bowen2015quantum}. 

Molecular polaritonics provides strikingly different tools to study and control biomolecular dynamics and reactions than have been available in the past. This offers a path towards a better understanding of complex biomolecular systems, allowing the roles of competing processes to be explored by tuning the chemical energy landscape.  For instance, it provides a new approach to explore how collective thermal vibrations of proteins drive functionally important conformational changes and thereby guide biological functions. This is important since, even though collective vibrations are thought to play a crucial role in processes ranging from energy transport to antibiotic resistance, protein folding, molecular binding and enzyme catalysis~\cite{schramm2018promoting,hay2012good,turton2014terahertz,niessen2019protein}, the mechanisms by which they do this are not clear~\cite{schramm2018promoting,niessen2019protein} and the concept itself remains controversial~\cite{schramm2018promoting,hay2012good,cao2020quantum}. Indeed, even quantum effects may play a role, as discussed in the following section. 
Molecular polaritonics also provides a route towards new technological applications, from higher yield chemical reactions and reactions that produce products that are otherwise inaccessible, to new quantum light sources for quantum computing and communications~\cite{o2009photonic} and quantum materials for efficient energy harvesting~\cite{gonzalez2015harvesting, zhong2017energy} built by exploiting robust room temperature biological systems.

\section{Quantum effects in biology}
\label{sec:QuantumEffectsBiology}

\begin{figure}[!ht]
 \centering
 \includegraphics[width=1\linewidth]{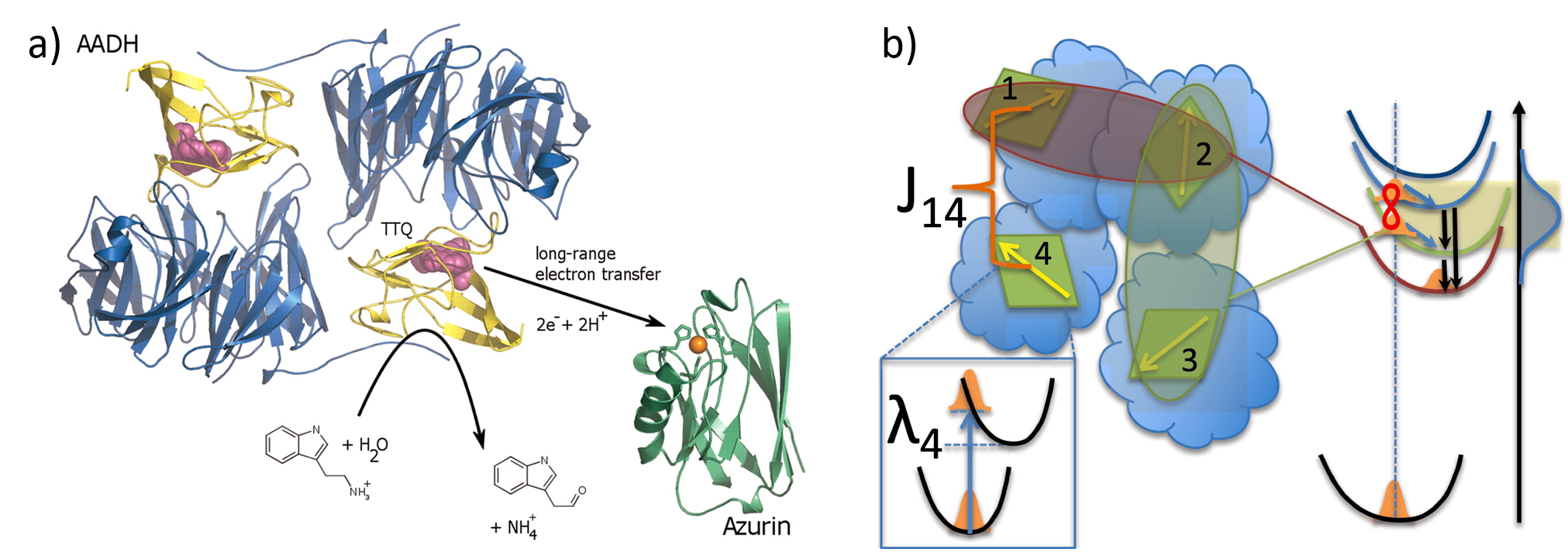}
 \caption{Quantum effects in biology. a) Enzyme reaction using quantum tunneling of proton. Proposed description of the AADH reaction with tryptamine (Reproduced from Ref.~\cite{masgrau2006atomic} with permission 5176190548612). b) Excitation energy transfer in photosynthesis. Left, green quadrilaterals represent chromophores that are situated in the protein environment (blue clouds). Right,  potential energy surfaces of the system and their transitions. (Reproduced from Ref.~\cite{manvcal2020decade} with permission 5176190756457)} 
\label{fig:quantumEffect}
\end{figure}

At the smallest scales of atoms and molecules all living systems follow the laws of quantum mechanics. For instance, electron orbitals and chemical bonds can only be properly understood from a quantum mechanical perspective. At large scales, however, it is expected that life obeys classical laws, without the need of deep quantum concepts such as superposition, coherence or entanglement. Where the boundary between classical and quantum descriptions exactly occurs, and whether biological systems have evolved to make functional use of quantum effects at larger scales than might otherwise be anticipated, has been an intense and controversial topic of debate since the birth of quantum mechanics. For instance, Erwin Schr\"{o}dinger argued for the importance of quantum mechanical for life in 1944 in his prescient book ``What is Life?''\cite{schro}. Other quantum mechanics luminaries have made similar arguments. Indeed, a lecture by Niels Bohr at the International Congress on Light Therapy in Copenhagen in 1933~\cite{bohr1933light} is credited with inspiring a young Max Delbr\"{u}ck who later contributed to founding the field of molecular biology and earned the 1969 Nobel Prize in Physiology or Medicine~\cite{isis,marais2018future}.

The field of quantum biology is fraught with controversies both because the definition of what exactly constitutes a quantum effect is not always agreed upon, and because it is highly non-trivial to unambiguously tease out these quantum effects from a complex biological background of other processes and activities~\cite{marais2018future, kim2021quantum}. A key argument is that non-trivial quantum effects should involve phenomena such as coherence, entanglement and superposition which one might usually expect to be averaged out by decoherence in warm, disordered biological environments, and that living system should exploit these effects for clear functional benefit. If such functionally relevant, non-trivial quantum effects do exist, this has the potential to transform  our understanding of life and to have major impact on the underpinnings of biotechnology, improving our understanding of phenomena such as the behaviour of biomolecules, the dynamics in neural network, biological energy transport and enzymatic catalysis (Fig.~\ref{fig:quantumEffect} a)). Better understanding of these processes is key to biotechnology applications ranging from the diagnosis and treatment of neurodegenerative disorders, to lower energy  fertilisers and biofuel production, more efficient energy technologies, and more effective pharmaceutical drug development. 

Quantum effects have been proposed to exist in processes ranging from olfaction~\cite{dyson1928some, turin1996spectroscopic,franco2011molecular}, to photosynthesis~\cite{engel2007evidence, lee2007coherence, manvcal2020decade}, enzyme catalysis~\cite{masgrau2006atomic, delgado2017convergence}, and even neural function~\cite{hameroff2014consciousness, hu2004spin, fisher2015quantum}. In each of these areas, significant knowledge gaps exist, that may be explainable via quantum effects. For instance, natural enzymes accelerate reactions by as much as fifteen orders of magnitude, but we do not know how to engineer artificial enzymes with the same capability~\cite{schramm2018promoting,hay2012good}.
It is known that quantum tunnelling plays a role in enzyme catalysis~\cite{masgrau2006atomic, delgado2017convergence} (Fig.~\ref{fig:quantumEffect} a)), though whether this is incidental or a key factor in explaining the performance of natural enzymes is yet to be determined. 
In photosynthesis  (Fig.~\ref{fig:quantumEffect} b)), we still do not fully understand the remarkable efficiency of photosynthetic energy transfer~\cite{cao2020quantum}.
Long-lasting coherence's in photosynthetic systems have been observed via two dimensional spectroscopy~\cite{engel2007evidence, lee2007coherence}. These were initially attributed to electron coherence and used as evidence that exciton transport to the photosynthetic reaction centre was accelerated by quantum interference between multiple pathways. It has since become clear that the situation is more complex, and that these coherences likely arise due to collective vibrational motions~\cite{duan2017nature, thyrhaug2018identification, jang2018delocalized}, which act as an ``environment'' that assists exciton transport~\cite{mohseni2008environment, higgins2021photosynthesis}. The robustness of environment-assisted  exciton transport in nature is under active investigation as a means to engineer high efficiency quantum transport \cite{shabani2014numerical, zerah2020effects}, with  quantum aspects predicted to improve transport at biologically relevant timescales and temperatures~\cite{o2014non}. 
In olfaction, it has been proposed that sensitivity to quantum mechanical collective vibrations is required to fully explain our sense of smell, augmenting lock-and-key mechanisms based on shape alone~\cite{dyson1928some, turin1996spectroscopic}. Experiments with fruit flies have provided evidence for this idea, by substituting hydrogen with deuterium in odorant molecules. This alters the vibrational frequencies of the odorant without significantly changing its shape, and was shown to result in a change in perceived odour~\cite{franco2011molecular}. Questions remain, however, with  experiments on humans and mouse odorant receptors finding no significant differences in perception for deuterated odorant molecules~\cite{block2015implausibility}. Given the exquisite ability of our sense of smell to differentiate between molecules, were quantum effects to prove significant they could enable the development of designer  molecular sensing tools with performance beyond what is currently possible.

Proposals that quantum effects occur in the brain are often  particularly controversial. Perhaps most notable is Roger Penrose and Stuart Hameroff's proposal that quantum coherence in neuronal microtubules may be capable of quantum computing and form the basis of human consciousness~\cite{hameroff2014consciousness}. This idea has not yet gained significant traction, with compelling arguments against coming from comparisons of the timescales for neuronal dynamics and for decoherence of the ions involved in the propagation
of action potentials~\cite{tegmark2000importance}. Motivated by the challenge of finding sufficiently low decoherence quantum states in the brain, and the suggestion that consciousness is linked to quantum spin~\cite{hu2004spin}, Matthew Fisher proposed the possibility of the spin of phosphorus nuclei as an alternative candidate for quantum neural processing~\cite{fisher2015quantum}. This is  attractive due to the long decoherence times of spin-half nuclear states, with hydrogen and phosphorous the only biologically relevant spin-half nuclei~\cite{fisher2015quantum}.

\begin{mdframed}
\vspace{4mm}
{\bf Box 1: Man versus machine: the remarkable aptitude of the human brain.}

While all proposals for quantum effects in the brain to date are of a speculative nature, they can be motivated by reminding ourselves of the seemingly unique feature of consciousness and of the remarkable power of the human brain. On the latter, consider the famous 2016 Go battle between Lee Sodol and Google's AlphaGo. Here, man and machine competed on equal footing in the most complicated game ever devised. While AlphaGo won four of five matches, the machine was designed specifically for Go (it could not ride a bicycle!) and, running on over a thousand central processing units and hundreds of graphical processing units, consumed around a megawatt of power compared to the light-bulb-level power consumption of the human brain. This is of course an anecdotal example, but is deeply suggestive that there is something powerfully different in how biological neural networks process information. Perhaps this involves quantum effects, perhaps not. If it does, the ramifications can be expected to be enormous. 

\vspace{4mm}

\end{mdframed}

\begin{figure}[!ht]
 \centering
 \includegraphics[width=1\linewidth]{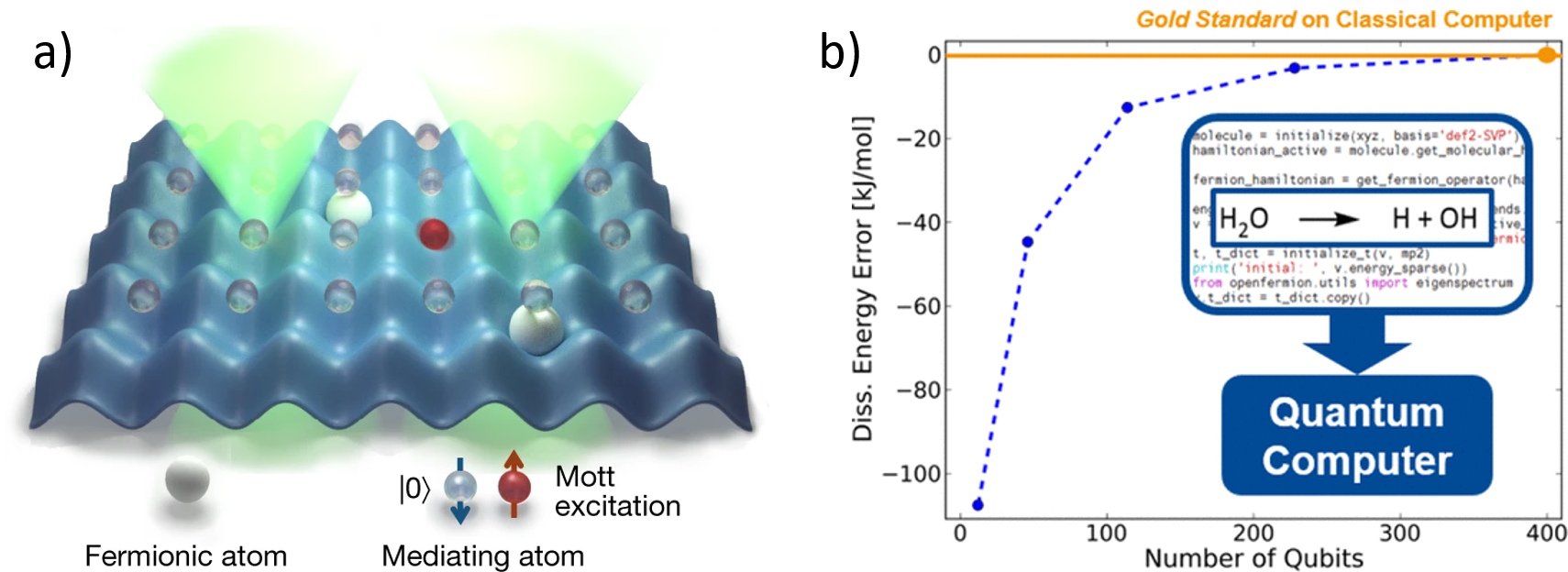}
 \caption{Quantum molecular simulation. a) Complete simulator of an analogue quantum chemistry simulation for the H2 molecule. (Reproduced from Ref.~\cite{arguello2019analogue} with permission 5176191038841) b) Quantum computer error versus number of qubits compared to the classical gold standard triple excitations coupled-cluster singles and doubles (CCSD(T)) (Reused with permission from Ref.~\cite{kuhn2019accuracy}. Copyright (2019) American Chemical Society).} 
\label{fig:quantumSimuation}
\end{figure}

Even after nearing a century of thought and exploration, the role of quantum effects in biology remains contentious. What has made the difference in recent years is the development of new tools, most especially two dimensional spectroscopy~\cite{tumbic2021protein}, that allow us to probe complex nanoscale biosystems and tweeze apart the different interactions and dynamics occurring within them. This has been complemented by advances in computational power and computational techniques to simulate molecular dynamics~(e.g. \cite{liguori2020molecular, gertig2020computer}). These technological developments can be expected to continue, with the sort of quantum tools to study and control biosystems with high precision discussed in the previous sections of  this review playing a key role. Beyond these new technologies, a further exciting prospect for progress in understanding the scope of quantum effects in biology is the rapid development of large-scale quantum computers~\cite{arute2019quantum}. While the range of practical problems that these quantum computers will be capable of addressing in the near future remains relatively limited , quantum simulations of molecular dynamics are one area they appear quite naturally suited to~\cite{mueck2015quantum, elfving2020will, kuhn2019accuracy,arguello2019analogue}. Here, they are able to address a fundamental problem in quantum chemistry calculations: the necessity to make approximations to avoid the exponential scale-up in complexity of computing the properties of quantum systems as their size increases~\cite{aaronson2009quantum}. The usual approach in quantum chemistry is to make the Born-Oppenheimer approximation, neglecting quantum correlations between electronic and nuclear degrees of freedom. While several notable methods exist to go beyond this approximation, they suffer from increased computational complexity, still require approximations, and ultimately are limited to relatively small molecules~\cite{aaronson2009quantum} (see Fig.~\ref{fig:quantumSimuation} a)). By leveraging quantum effect themselves, quantum computers can avoid approximations without an exponential increase in simulation time. This provides the prospect to better understand both how quantum effects influence the behaviour of biomolecules, and how to engineer these effects in artificial molecules. Quantum computers have already be used to simulate a range of small molecules~(see summary in \cite{elfving2020will}), but larger, more fault-tolerant computers are expected to be required to simulate molecules relevant to biotechnology~\cite{elfving2020will,kuhn2019accuracy} (see Fig.~\ref{fig:quantumSimuation} b)). It has been predicted, for example, that in the medium term quantum computers could simulate inorganic catalysts such as those used to fix nitrogen or convert
carbon dioxide into methanol~\cite{elfving2020will,kuhn2019accuracy}. This could assist us in better understanding these critical biotechnological processes.

\section{Conclusion}

This review paper  provides an overview of recent developments at the interface of quantum and biological science, and an outlook to their possible future impacts on biotechnology. These developments span a wide breadth of areas, from better sensors and microscopes, to new methods to simulate biochemical dynamics, to quantum control of biomolecules and chemical reactions, and even to the possible presence of quantum effects evolved to drive functional behaviours within biological systems themselves. The possible impact of these quantum technologies includes the development of a better understanding of the biodynamics that underpin pharmaceutical development, enzyme function and photosynthetic energy transport; new ways to improve the yield of chemical reactions or to produce new chemical species;  advanced imaging technologies that can be used to diagnose disease and understand the pathways by which drugs act; and answers to deep questions on the role of quantum effects  in living systems.  This paints an exciting picture of a future where quantum technologies drive new advances in biotechnology.

% Acknowledgements
\medskip
\textbf{Acknowledgements} \par %delete if not applicable))
This material is based upon work supported by the Air Force Office of Scientific Research under award
number FA9550-20-1-0391. It was also supported by the
Australian Research Council Centre of Excellence for Engineered Quantum Systems (EQUS, CE170100009). 

\medskip

\textbf{Conﬂict of Interest}

The authors declare no conﬂict of interest.

% References
\medskip

% Use the following code if you wish to generate your bibliography with BibTeX;
% replace the string "MSP-template" below with the name(s) of
% the BibTeX data base(s) you want to use.
% The resulting bibliography-output (the content of the .bbl file)
% must be pasted back into this file before submission.
% Please also include your BibTeX data base file(s) in your submission
% so that we can re-run BibTeX if necessary.

% \bibliographystyle{MSP}
% \bibliography{refs}

% Please provide Biographies and photos for Essays, Feature Articles, Progress Reports, Reviews, and Perspectives for those authors who should be highlighted  
% These should be at most 100 words long
% For other article types this section can be removed
% Photographs should be 40mm broad and 50 mm high
\newpage

\begin{figure}
  \includegraphics{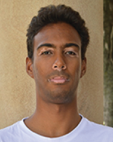}
  \caption*{\textbf{Nicolas P. Mauranyapin} received his M.Sc. degree in 2011 from the university of Strasbourg (France) and his Ph.D. degree in 2018 from the University of Queensland (Australia). He is currently a Postdoctorate-Fellow at the Queensland quantum optics lab and his research interests focus on  translation of quantum technology to biotechnology and nano-mechanical computing.}
\end{figure}

\begin{figure}
  \includegraphics{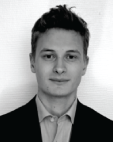}
  \caption*{\textbf{Alex Terrason} received his master's degree from the university of Lyon I (France) in 2018. He started his Ph.D candidature at the University of Queensland (Brisbane, Australia) in 2019. His research focuses on quantum optical microscopy and the study of Brownian motion in living, out-of-equilibrium systems.}
\end{figure}

\begin{figure}
  \includegraphics{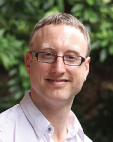}
  \caption*{\textbf{Professor Warwick P. Bowen}’s research focuses on the implications of quantum science on precision measurement, and applications of quantum measurement in areas ranging from quantum condensed matter physics to the biosciences. He is a Fellow of the Australian Institute of Physics, Director of the University of Queensland Precision Technologies Translation Hub, and a Theme Leader of the Australian Centre for Engineered Quantum Systems. His lab has significant efforts in using quantum light and quantum-limited technologies to improve biological microscopy. They also have active research efforts on integrated photonics, quantum control of macroscopic mechanical devices, and superfluid helium physics.}
\end{figure}

\end{document}